\documentclass[reprint,amsmath,amssymb,aps]{revtex4-2}
\usepackage{graphicx} 
\usepackage{dcolumn}
\usepackage{bm}
\usepackage[colorlinks=true, citecolor=red, linkcolor=blue, urlcolor=blue ]{hyperref}
\newcommand{\red}[1]{\textcolor{red}{#1}}  
\usepackage[T1]{fontenc}

\begin{document}

\title{The Architecture of Sponge Choanocyte Chambers Maximizes\\ Mechanical Pumping Efficiency}
\author{Takumi Ogawa}
\email[]{takumi.ogawa.t2@dc.tohoku.ac.jp}
\affiliation{Department of Finemechanics, Tohoku University, 6-6-01 Aoba, Aramaki, Aoba-ku, Sendai 980-8579, Japan}
\author{Shuji Koyama}
\affiliation{Department of Finemechanics, Tohoku University, 6-6-01 Aoba, Aramaki, Aoba-ku, Sendai 980-8579, Japan}
\author{Toshihiro Omori}
\email[]{omori@tohoku.ac.jp}
\affiliation{Department of Finemechanics, Tohoku University, 6-6-01 Aoba, Aramaki, Aoba-ku, Sendai 980-8579, Japan}
\author{Kenji Kikuchi}
\email[]{k.kikuchi@tohoku.ac.jp}
\affiliation{Department of Finemechanics, Tohoku University, 6-6-01 Aoba, Aramaki, Aoba-ku, Sendai 980-8579, Japan}
\author{H{\'e}l{\`e}ne de Maleprade}
\email[]{helene.de\_maleprade@sorbonne-universite.fr}
\affiliation{Institut Jean Le Rond d'Alembert, Sorbonne Université, CNRS UMR 7190, 75005 Paris, France}
\author{Raymond E. Goldstein}
\email[]{R.E.Goldstein@damtp.cam.ac.uk}
\affiliation{Department of Applied Mathematics and Theoretical Physics, University of Cambridge, Wilberforce Road, Cambridge, CB3 0WA, United Kingdom}
\author{Takuji Ishikawa}
\email[]{t.ishikawa@tohoku.ac.jp}
\affiliation{Department of Biomedical Engineering, Tohoku University, 6-6-01 Aoba, Aramaki, Aoba-ku, Sendai 980-8579, Japan}

\date{\today}

\begin{abstract}
Sponges, the basalmost members of the animal kingdom, exhibit a range of complex 
architectures in which 
microfluidic channels connect multitudes of spherical chambers lined with choanocytes, 
flagellated filter-feeding cells.
Choanocyte chambers can possess scores or even hundreds of such cells, which drive complex 
flows entering through porous 
walls and exiting into the sponge channels.
One of the mysteries of the choanocyte chamber is its spherical shape, as it seems 
inappropriate for inducing directional 
transport since many choanocyte flagella beat in opposition to such a flow.  Here 
we combine direct imaging 
of choanocyte chambers in living sponges with computational studies of many-flagella 
models to understand 
the connection between chamber architecture and directional flow.  We find that those 
flagella that beat against 
the flow play a key role in raising the pressure inside the choanocyte chamber, with 
the result that the mechanical 
pumping efficiency, calculated from the pressure rise and flow rate, reaches a maximum 
at a small outlet opening angle.
Comparison between experimental observations and the results of numerical simulations 
reveal that the chamber diameter, 
flagellar wave number and the outlet opening angle of the freshwater sponge 
\textit{E. muelleri}, as well as several 
other species, are related in a manner that maximizes the mechanical pumping efficiency.
These results indicate the subtle balances at play during morphogenesis of choanocyte chambers, and give insights into the physiology and body design of sponges.
\end{abstract}

\maketitle

\section{Introduction}
\begin{figure*}
\includegraphics[width=1.75\columnwidth]{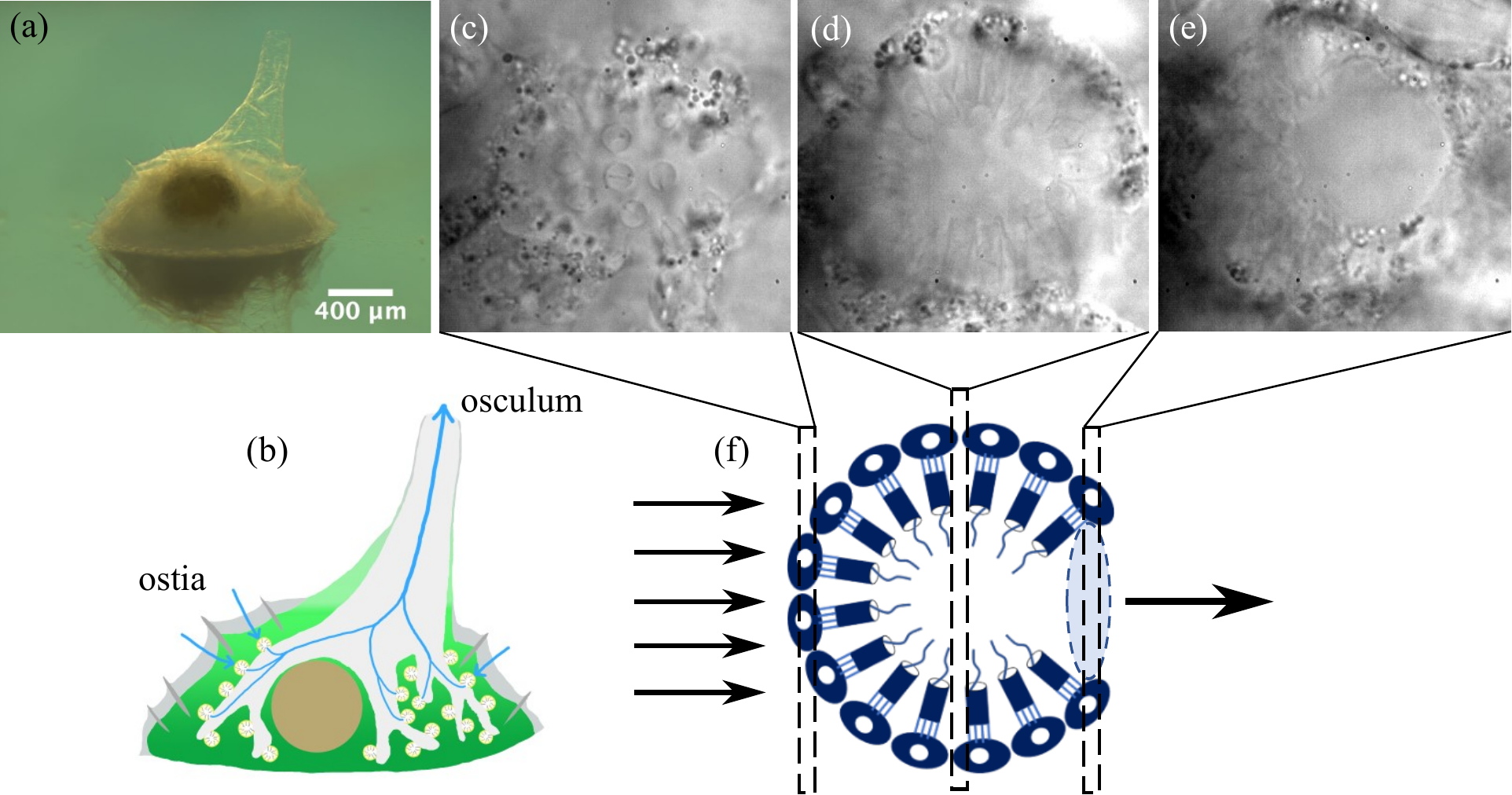}
\caption{The fresh water 
sponge \textit{E. muelleri}. (a) Stereomicroscopic image, with the body shown reflected in glass. (b) Schematic diagram of the aquiferous canal system. Blue lines indicate the direction of water flow.
(c-e) Microscopic images of the choanocyte chamber at the bottom, middle and top focal planes as indicated in the
schematic (f).  Parallel arrows indicate fluid flow into a choanocyte chamber, while blue shaded ellipse indicates location of the opening from which fluid exits.}
\label{fig1}
\end{figure*}

Sponges are among the oldest multicellular animals, with 
sponge-like fossils having been found that date as far back as 550-760 million years \cite{Brain2012}. 
As filter feeders with a distinct pump-filter apparatus, sponges (phylum Porifera) 
can process several hundred times their body volume of 
water per hour \cite{Reiswig1971,Kahn2015}.
Because of this, they play a significant role in nutrient cycling within marine ecosystems 
such as coral reefs \cite{Goeij2008,Goeij2013,Achlatis2019}.
To achieve this high-performance filtering, sponges have an evolved aquiferous system 
that allows water to flow through their bodies \cite{Hammel2012,Yin2015}, and
are divided into several classes (termed {\it asconoid}, {\it syconoid}, and {\it leuconoid}) according to the level of complexity of this internal microfluidic system \cite{spongetextbook}.
Leuconoid sponges have the most complex architecture, composed of incurrent and excurrent canals. 
Water entering their body through the inlets (ostia) on the surface of the sponge body is carried through the incurrent/excurrent canals and exits from the outlet (osculum), as shown in Fig.~\ref{fig1} 
for the case of the freshwater sponge \textit{Ephydatia muelleri}.
The spherical chambers act as pumps between the incurrent and excurrent canals, and are lined with 
flagellated cells termed choanocytes arranged in a radial pattern with flagella directed toward the center 
of the sphere, supporting traveling waves of motion that direct flow toward the center.  
Filtering of incoming water is achieved by a collar of microvilli anchored near 
the apex from which emanates the flagellum of each cell.

Both phylogenomics \cite{Redmond2021,Simion2017,Telford2016} and morphology \cite{Nielsen2008,Nielsen2019b} support 
the ``Porifera-first" theory of animal evolution. While some studies suggest that the common ancestor of Metazoa appeared 
from a unicellular choanoflagellate ancestor, because choanoflagellates are closely related to Metazoa and are 
morphologically similar to sponge choanocytes \cite{Nielsen2008,Laundon2019,Pinskey2022}\textemdash 
it has been confirmed from a molecular phylogenetic view that the two groups are monophyletic \cite{Carr2008} and
it is difficult to assume homology between these two cells due to functional differences \cite{Mah2014}.
Thus, while the analysis of the evolution from unicellular to multicellular life remains unsettled, it remains likely that sponges are the oldest animals, and their simple body plans make them a model organism for studies of both morphogenesis and physiology.

The spherical arrangement of choanocytes (Fig.~\ref{fig1}(c-f) appears at first glance to be ill-suited to yield 
directional flow through the chamber, since inevitably the flagella of some of the choanocytes would beat in 
opposition to the flow.  Since sponges have survived from the distant past, it is natural 
to hypothesize that this spherical shape confers an evolutionary benefit.  
Here we seek to understand this issue 
from the perspective of fluid mechanics.

\begin{figure*}[t]
\includegraphics[width=1.85\columnwidth]{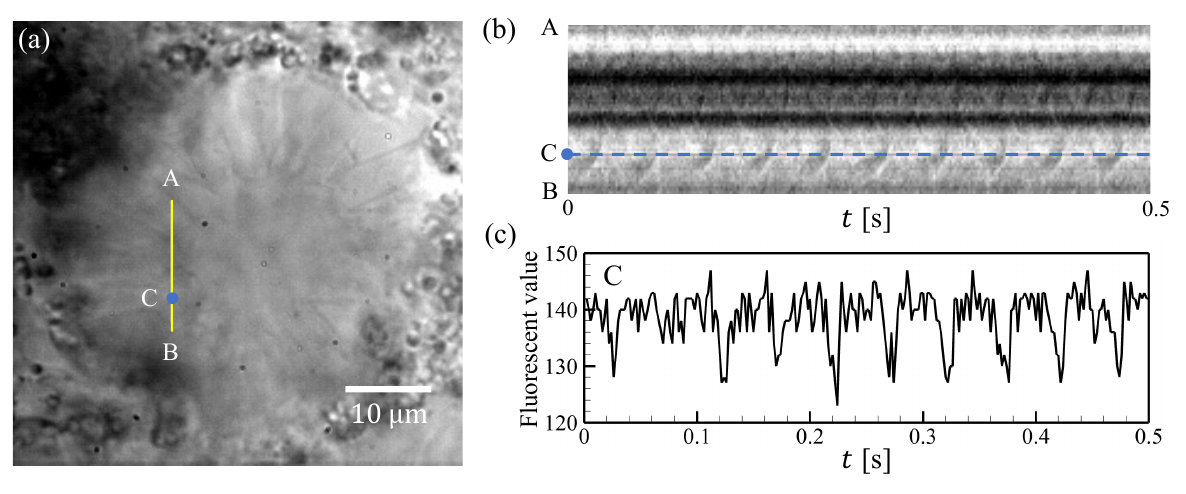}
\caption{Flagellar beating in a choanocyte chamber of \textit{E. muelleri}. 
(a) Microscope image of a choanocyte chamber. (b) Spaciotemporal brightness 
fluctuation in the choanocyte chamber showing flagellar beating. (c) Slice 
of data in (b) showing the temporal oscillations in pixel intensity.}
\label{fig2}
\end{figure*}

Vogel’s pioneering early work was the first to highlight the fluid mechanics of 
sponges \cite{Vogel1974,Vogel1977,Vogel1978}.   While the properties of the ``sponge pump" have been of 
interest since then \cite{Larsen1994,Riisgard1995}, they remain incompletely 
understood \cite{Ludeman2017,Hammel2014,Goldstein2019}.  One of the most important issues is the 
connection between large-scale flows external to the sponge and the flows within.  Although 
the sizes of sponges span an enormous range, even taking a modest vertical scale $L\sim 10-10^2\,$cm 
and the typical ambient flow speeds $U\sim 10-10^2\,$cm/s the Reynolds number $Re=UL/\nu$ 
in water is $10^4-10^5$ 
outside a sponge.   With such a large $Re$ we expect a viscous boundary to form at the sponge surface whose
maximum thickness $\delta\sim x/\sqrt{Re_x}$, where $x$ is distance along the sponge and $Re_x=Ux/\nu$.  Using
the representative $x\sim 10\,$cm we obtain $\delta\sim 1\,$mm.  Thus, the ostia through which water flows into the
microfluidic network of the sponge sit within the viscous boundary layer.  
While the general problem of flow into a
permeable wall has been studied in some detail \cite{Beavers1967,Vafai1990,Fransson2004}, 
no connection has yet been made with sponge fluid dynamics. 

The relationship between external and internal flows can then be recast in terms of the connection between 
Bernoulli pressure differences between
the ostia and osculum set up by external flows and internal pressures created by choanocytes beating within the
chambers.  
In this regard there are conflicting conclusions in the literature.
Dahihande and Thakur \cite{Dahihande2021} reported that the number density of choanocyte chambers and 
the number of choanocytes per chamber are positively correlated with the pumping rate of sponges. On the 
other hand, Larsen and Riisgård \cite{Larsen2021} found that the pumping rate depends on the pressure losses 
of the aquiferous system with increasing sponge size and not on the reduced density of choanocytes.
Recently, there have been several important numerical studies of the sponge pump 
at the level of choanocytes. Asadzadeh et al. \cite{Asadzadeh2019} found numerically that the both 
the glycocalyx mesh covering the upper part of the collar and secondary reticulum are important for 
the pump to deliver high pressure. They also reported that choanocytes arranged in a cylindrical 
configuration can pump water efficiently owing to the formation of a hydrodynamic gasket 
above the collars \cite{Asadzadeh2020}. However, the spherical shape of choanocyte chambers was not discussed 
hydrodynamically in any of these former studies and the physiological significance of that shape is unclear.

Here we examine the biological significance of the body design of spherical choanocyte chambers by combining 
direct imaging of choanocyte chambers in living sponges with computational studies of many-flagellum models of 
their fluid mechanics. We used freshwater sponges \textit{Ephydatia muelleri} as a model organism for experimental 
observation and to define a computational model of the fluid mechanics of a choanocyte chamber. 
The experimental and computational 
methodologies are explained in Sections \ref{ExpMethods} and \ref{CompMethods}.
In Section \ref{Results}, we show computationally that the flagella beating against the flow play a role 
in raising the pressure inside the choanocyte chamber and the mechanical pumping efficiency reaches 
a maximum at a certain outlet opening angle. Section \ref{Comparison} shows the 
good agreement between the experimental and the numerical results.
Section \ref{Force_model} investigates the extent to which the flows in a 
choanocyte chamber can be represented by a small number of Stokes singularities.
In the summary Section \ref{Conclusions}, 
we suggest further avenues of research on sponge physiology and development.

\section{Experimental Methods}
\label{ExpMethods}

\textit{Harvesting and culturing of sponges:}
Freshwater sponges of the species \textit{Ephydatia muelleri}, shown in Fig.~\ref{fig1}(a), living 
on natural stones were collected 
from the Hirose River (Izumi, Sendai City, Miyagi, Japan). The asexually produced buds of reproductive 
cells known as gemmules were peeled off, and the spicules on them were removed. To remove impurities, 
the gemmules were washed with a $1$\% aqueous solution of H$_2$O$_2$, followed by three successive rinses
with pure water to remove excess H$_2$O$_2$. The purified gemmules were stored in a refrigerator at 4°C. 
Sponges were cultured in a plastic dish with Strekal's medium at room 
temperature ($25^\circ$C).  

\textit{Imaging:}
Choanocyte chambers of \textit{E. muelleri} were imaged on an inverted microscope 
(IX71, Olympus Corp., Tokyo, Japan), as shown in Fig.~\ref{fig2}(a).  Flagellar beating 
in the chambers was imaged with a oil immersion objective (100$\times$, N.A.= 1.40) 
and a high speed camera (500 fps, 1024$\times$1024 pixels, FASTCAM SA3, Photron 
Limited, Tokyo, Japan) for durations
of $5\,$s. The beating appears as a spatiotemporal brightness fluctuation in the 
pixels of the images \cite{Kikuchi2017,Omori2022}, as shown in Fig.~\ref{fig2}(b),
from which the beat frequency was measured from a fast Fourier transform (n=5, N=3).  The various geometric and
dynamical characteristics of choanocyte chambers (n=6-8, N=4) and the flagellar motion within them (n=3-5, N=4) are shown in 
Table~\ref{table1}, in which the results of \textit{Spongilla lacustris} obtained by Mah et al. 
\cite{Mah2014} are added for comparison.

\begin{table}[b]
\caption{Characteristics of choanocyte chambers.}
\begin{ruledtabular}
\begin{tabular}{llll}
Quantity & \textit{E. muelleri} & \textit{S. lacustris}\footnote{Data for \textit{Spongilla lacustris} 
are from Mah et al. \cite{Mah2014}.}& Simulation\\
\hline 
number of flagella $N$ & 112$\pm$31 & & 37-359\\
flagellar length $L$ & 12.9$\pm$1.1 $\mu$m & 10.4$\pm$0.3 $\mu$m & 13 $\mu$m \\
flagellar amplitude $A$ & 1.58$\pm$0.21 $\mu$m & 3.3$\pm$0.9 $\mu$m & 0.14$L$\\
beat frequency $\nu$ & 26.1$\pm$8.9 Hz & 11.0$\pm$1.1 Hz & \\
chamber radius $R$ & 17.4$\pm$2.5 $\mu$m & & (1.5-3)$L$\\
\end{tabular}
\end{ruledtabular}
\label{table1}
\end{table}

\begin{figure*}
\includegraphics[width=1.85\columnwidth]{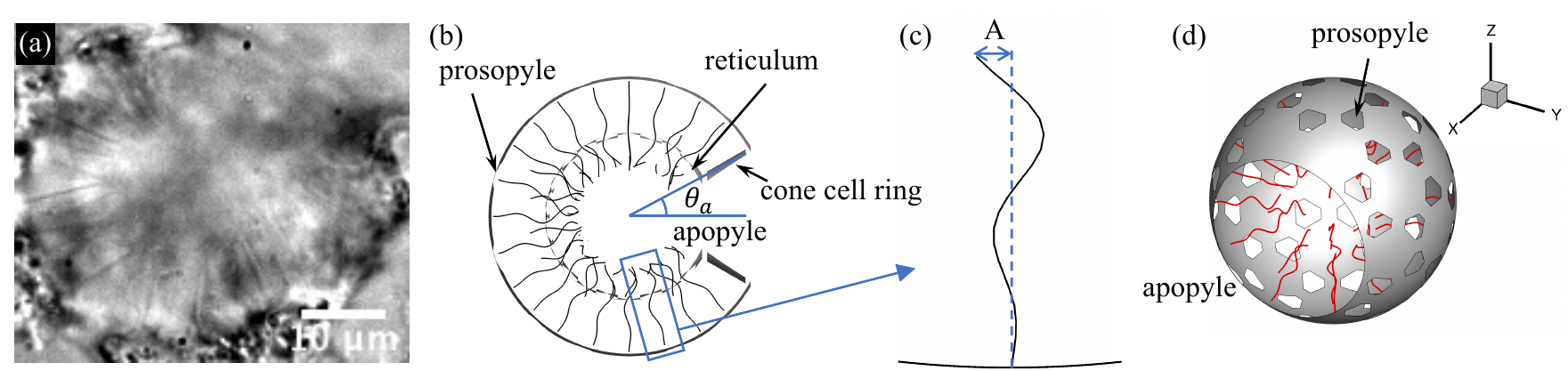}
\caption{Computational model of choanocyte chamber. (a) A choanocyte chamber of \textit{E. muelleri}. (b) Schematic cross section of the model 
chamber. (c) Definition of the flagellar beat amplitude $A$. 
(d) Three-dimensional model of the choanocyte chamber.}
    \label{fig3}
\end{figure*}

\section{Computational Fluid Dynamics of choanocyte chambers}
\label{CompMethods}
Choanocyte chambers consist of choanocytes arranged in a spherical configuration, 
each with a single flagellum oriented toward the center of the chamber, propagating
bending waves radially inward.  In the computations, we represent
the chamber as a rigid sphere of radius $R$ from which flagella emerge, as in Fig.~\ref{fig3}. 
The chamber has many small inlet 
holes termed {\it prosopyles} and one large outlet hole known as the {\it apopyle}. 
Additionally, there is the gasket-like {\it reticulum}, 
a fine structure that connects the apical part of the collars, which was also 
observed in \textit{Spongilla lacustris} \cite{Weissenfels1992}.
The reticulum delivers high pressure because the choanocyte acts as a 
collar-vane-flagellum pump system \cite{Asadzadeh2019}.
The reticulum was modeled as a concentric spherical shell inside the choanocyte chamber 
(Fig.~\ref{fig3}(b)), and its
radius was varied with the chamber radius so that the length of 
the collar extending from the choanocyte was a constant $8.2\,\mu$m \cite{Mah2014}.
Finally, {\it cone cells} are found in several freshwater sponges including 
\textit{E. muelleri} \cite{Langenbruch1987}, and are flattened near 
apopyles \cite{Weissenfels1992,Langenbruch1983,Langenbruch1987}. 
They form a ring and connect collars with the walls of excurrent 
canals to prevent the disadvantageous current.
The cone cell ring was modeled as a conical surface near the apopyle (Fig.~\ref{fig3}(b)), 
with an opening angle $\theta_a$.

\textit{Governing equations:}
Scaled by the flagellar motion, the Reynolds number $Re\ll 1$ in a
chamber, so  
the fluid flow in and around choanocyte chambers is governed by the Stokes equations, and
we use slender-body theory \cite{Andersson2021} for the flagella.
Their centerlines are parameterized by arclength $s \in [0,L]$,  
and we measure the chamber radius $R$ in units of the flagellar length, with
\begin{equation}
    \rho=\frac{R}{L}
\end{equation}
a parameter that controls the size of the central region of the chamber devoid of flagella.
The velocity $\bm{v}$ at point $\bm{x}\in s_i$ located on flagellum $i$ can be written as \cite{Ito2019}
\begin{align}
    \bm{v}(\bm{x})=&-\frac{1}{8\pi\mu}\int_{ch} \bm{J}(\bm{x},\bm{y})\cdot \bm{q}(\bm{y}) dS(\bm{y}) \nonumber\\
    &-\frac{1}{8\pi\mu}\sum_{j}^{N}\int_{fla} \bm{K}(\bm{x},\bm{y}) \cdot \bm{f}(\bm{y}) ds_j(\bm{y}),
\label{eq:fla}
\end{align}
where $\mu$ is the viscosity, $\bm{q}$ is the viscous traction on the chamber, $\bm{f}$ is the force density of the flagella and $N$ is the total number of flagella. 
The first integral is over all surfaces $S$, including those of
the chamber, reticulum and cone cell ring.
The second integral is over flagellar centerlines. In \eqref{eq:fla},
$\bm{J}$ is the Green's function
\begin{eqnarray}
    J_{ij}(\bm{x},\bm{y})=\frac{\delta_{ij}}{r}+\frac{r_i r_j}{r^3},
\end{eqnarray}
where $r=\vert \bm{r} \vert$ and $\bm{r}=\bm{x}-\bm{y}$, and
$\bm{K}$ is the kernel \cite{Andersson2021}
\begin{equation}
     K_{ij}=\frac{\delta_{ij}}{b} + \frac{r_i r_j}{b^3} 
     + \frac{{\varepsilon}^2 L^2 a^2(s')}{2} \left( \frac{\delta_{ij}}{b^3} - \frac{r_i r_j}{b^5} \right), 
\end{equation}
where $b=\sqrt{r^2 + \varepsilon^2 a^2(s)}$.
Here the radius function $a$ satisfies $0 < a(s) \leq 1$ for each $s \in [0, \sqrt{L^2+\varepsilon^2}]$, where $\varepsilon$ is the ratio of the length to the radius of flagellum and is set to $\varepsilon = 0.01$.
The radius function $a(s)$ is 
\begin{eqnarray}
    a(s) = \frac{1}{\sqrt{L^2 + \varepsilon^2}} \sqrt{L^2 + \varepsilon^2 - s^2}.
\end{eqnarray}
When the point $\bm{x}$ is not on a flagellum, the velocity is 
\begin{align}
    \bm{v}(\bm{x})=&-\frac{1}{8\pi\mu}\int_{ch} \bm{J}(\bm{x,y})\cdot \bm{q}(\bm{y}) dS(\bm{y}) \nonumber\\
    &-\frac{1}{8\pi\mu}\sum_{j}^{N}\int_{fla} 
    \left[\bm{J}(\bm{x},\bm{y}) + \bm{V}(\bm{x},\bm{y})\right] \cdot \bm{f}(\bm{y}) ds_j(\bm{y}), \nonumber\\
\label{eq:v}
\end{align}
where \cite{Andersson2021,Tornberg2004}
\begin{eqnarray}
     V_{ij} = \frac{{\varepsilon}^2 L^2 a^2(s')}{2} \left( \frac{\delta_{ij}}{r^3} - \frac{r_i r_j}{r^5} \right).
\end{eqnarray}

\begin{figure*}
\includegraphics[width=1.75\columnwidth]{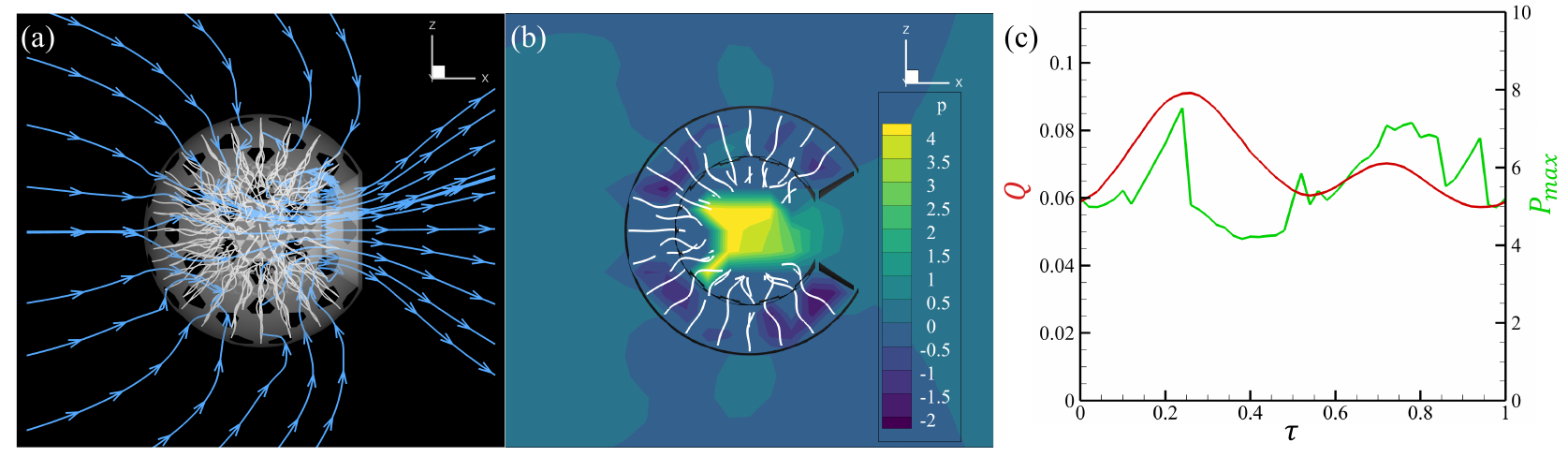}
\caption{Numerical results. (a) Flow field around the model choanocyte chamber, 
with $\rho=1.5$, $N=143$ ($\phi=0.33$), $\theta_a=30^\circ$ and $k\ell=3\pi$. 
The streamlines are drawn in the center cross-section. 
(b) Pressure distribution in the chamber. (c) Temporal behavior of the outlet flow rate 
and maximum pressure within the chamber.}
    \label{fig:flow}
\end{figure*}

\textit{Flagellar motions:}
Each of the flagella within a computational choanocyte chamber is actuated with a prescribed waveform
at a frequency $\nu=1/T$, where $T$ is the period of motion,
and fixed at its base on the choanocyte chamber surface at a point $\bm{x}_b$.
We define the orthonormal body frame $\bm{g}_1$ and $\bm{g}_2$ at $\bm{x}_b$ as $\bm{g}_1(\bm{x}_b)=-\bm{n}(\bm{x}_b)$, and $\bm{g}_2(\bm{x}_b) = \bm{b}(\bm{x}_b) \wedge \bm{n}(\bm{x}_b)/|\bm{b}(\bm{x}_b) \wedge \bm{n}(\bm{x}_b)|$, where $\bm{b}(\bm{x}_b) = \bm{e}_1 \wedge \bm{n}(\bm{x}_b)$, $\bm{e}_1 = (1,0,0)$ and $\bm{n}(\bm{x}_b)$ is the outward unit normal vector to the surface.
With $\tau=t/T$, the motion of a flagellum is parameterized as
\begin{equation}
    \bm{x}^{fla} = \xi \bm{g}_1(\bm{x}_b) +  \frac{A \xi}{\ell (t)} 
    \cos \left(\frac{kL\xi}{\ell (t)}
    - 2\pi \tau + \varphi \right)\bm{g}_2(\bm{x}_b),
    \label{eq:flagmotion}
\end{equation}
where $A$ is the beat amplitude, $k=2\pi/\lambda$, with $\lambda$ the wavelength, the 
coordinate $\xi \in[0, \ell (t)]$ spans the time-dependent {\it projected} length 
$\ell(t)$ of the flagellum under the constraint of fixed total arclength $L$.
For the oscillating flagellum described by $\eqref{eq:flagmotion}$,
the projected arclength $\ell$ is considerably less than the total arclength $L$.  
For the amplitude $A=0.14L$ used in numerics and for the values of $k\ell\sim (3-4)\pi$ typical of
experiment we have $\ell/L\sim 0.76$.  This contraction plays an important role in the pressure
distribution within the choanocyte chamber.
As we observed no phase synchrony in our studies of sponge flagella, much as
earlier studies of multicellular choanoflagellates saw no synchrony 
\cite{Roper2013,Kirkegaard2016}, in computations 
we randomly set the phase $\varphi$
for each flagellum in the range $\varphi \in[0,2\pi]$,
reproduce the independent flagellar motion in the chamber.

\textit{Boundary element method:}
When the choanocyte chamber is fixed and there is no background flow, the boundary conditions are
\begin{eqnarray}
  \bm{v}(\bm{x}) = 
  \begin{cases}
        0 &\text{$\bm{x} \in$ chamber} \\
        \bm{v}^{fla} (\bm{x}) &\text{$\bm{x} \in$ flagella}
  \end{cases}
\label{boundary}
\end{eqnarray}
where $\bm{v}^{fla} = {\partial \bm{x}^{fla}}/{\partial t}$ is the flagella velocity, and
the chamber includes the reticulum and cone cell ring.

The choanocyte chamber surface, reticulum and cone cell ring were composed of 9,710 triangular 
mesh elements and 5,401 nodal points. These quantities depended on the chamber radius, the size of apopyles, and the number of flagella.
Each flagellum was discretized into 20 elements with 21 nodes.
All physical quantities were computed at each discretized point.
The boundary integral equations \eqref{eq:fla}) and \eqref{eq:v} were computed using 
Gaussian integration, leading to the linear algebraic equations
\begin{align}
    [\bm{v}^c] &= [J^{cc}][\bm{q}] + [J^{cf}][\bm{f}],\\
    [\bm{v}^f] &= [J^{fc}][\bm{q}] + [J^{ff}][\bm{f}].
\end{align}

Both $[\bm{v}^c]$ and $[\bm{q}]$ have size $3N_c$, and that of 
$[\bm{v}^f]$ and $[\bm{f}]$ is $3N_f$, where $N_c$ is the total number of nodes on 
the chamber surface, the reticulum and the cone cell ring, and $N_f$ is the total 
number of nodes on the flagella.
The matrix size of $[J^{cc}]$ is $3N_c \times 3N_c$ and $[J^{cf}]$ is $3N_c \times 3N_f$.
Considering the boundary condition \eqref{boundary}, the equations can be rewritten as 
\begin{eqnarray}
    \begin{bmatrix}
        J^{cc} & J^{cf} \\
        J^{fc} & J^{ff}
\end{bmatrix}
\begin{bmatrix}
        \bm{q} \\
        \bm{f}
\end{bmatrix}
=
\begin{bmatrix}
        0 \\
        \bm{v}^{fla}
\end{bmatrix}.
\end{eqnarray}
The dense matrix was solved using the lower-upper (LU) factorization technique.

\textit{Parameters:}
To investigate the effects of the choanocyte chamber geometry on the pumping 
function of the chamber, in the numerical studies reported below we studied 
a range of several of the 
parameters of the computational modes: number of flagella $N\in[37-359]$, 
chamber radius $R\in [1.5L,3L]$ (with $L=13 \mathrm{\mu m}$), scaled
wavenumber $k\ell\in[1.5\pi,3\pi]$, and apopyle opening 
angle $\theta_a\in[10^\circ,90^\circ]$.
Through all computations, the calculational time step $\Delta t$ was set to $0.02 T$, where $T$ is the flagellar beat period; we also used this in the time averaging method discussed below.

\textit{Choanocyte packing fraction:}  In describing the numerical computations, the various parameters of
the setup can be recast in convenient dimensionless quantities.  
First, if $\theta_a$ is the opening angle of the apopyle, the area of
the chamber available for choanocytes is $2\pi R^2(1+\cos\theta_a)$.  We may view the flagellar undulation
amplitude $A$ in \eqref{eq:flagmotion} as defining an area $\pi A^2$ associated with each choanocyte, and thus
it is natural to define the area fraction $\phi$ of choanocytes on the chamber surface as
\begin{equation}
    \phi=\frac{NA^2}{2R^2\left(1+\cos\theta_a\right)}.
\end{equation}
For reference, in a hexagonal close packing on a flat surface, the maximum packing fraction is 
$\pi\sqrt{3}/6\simeq 0.907$.

\section{Numerical Results}
\label{Results}

\subsection{Basic observations}

\textit{Flow field:}
\label{flow_field}
The time-averaged flow field and pressure in the center cross-section of a chamber are 
shown in Figs.~\ref{fig:flow}(a) and (b) for representative parameters.
We confirmed that the unidirectional flow, from prosopyle to apopyle, is generated by 
flagellar beating in the spherical chamber.
In this spherical geometry, the pressure field can be divided into low and high pressure regions, 
as previously reported \cite{Weissenfels1992,Asadzadeh2019}.
The low pressure region sucks water from the prosopyle, while the high pressure region 
ejects water outward, leading to a unidirectional flow with
volumetric rate $Q^*$ from the apopyle of
\begin{equation}
    Q^* = \int\! \bm{u} \cdot \bm{n} \,dS_a,
\label{eq:Q}
\end{equation}
where $S_a$ is the area of the curved surface of the apopyle, $\bm{u}$ is the 
velocity on $S_a$, and $\bm{n}$ is the outward unit normal vector on $S_a$. 
As the flagellar length is the fixed scale used for nondimensionalizing quantities, we use it
and the beat period to define the rescaled flow rate
\begin{equation}
    Q=\frac{Q^*T}{L^3}
\end{equation}
and its time-averaged value $\bar{Q}$.
As shown in Fig.~\ref{fig:flow}(c), $Q$ does not vary significantly with time. 
For the parameters $\rho=1.5$, $N=143$ ($\phi=0.33$), $\theta_a=30^\circ$ and 
$k\ell=3\pi$ we find $\bar{Q}=0.071$, a value that is  
compared with experimental results in Section \ref{Comp_flow}.

The pressure at point $\bm{x}$ is \cite{Pozrikidis1992,Lac2007}
\begin{align}
    P^*(\bm{x})=&-\frac{1}{8\pi\mu}\int_{ch} \bm{p}(\bm{x},\bm{y})\cdot \bm{q}(\bm{y}) dS(\bm{y}) \nonumber\\
    &-\frac{1}{8\pi\mu}\sum_{i}^{N}\int_{fla} \bm{p}(\bm{x},\bm{y}) \cdot \bm{f}(\bm{y}) ds_i(\bm{y}),
\label{eq:P}
\end{align}
where $\bm{p}=2\bm{r}/r^3$.  From the Stokesian balance $\bm{\nabla}P^*=\mu\nabla^2{\bf u}$, with
$u\sim L/T$, a suitable rescaled pressure is
\begin{equation}
    P=\frac{P^*T}{\mu},
\end{equation}
which is used to quantify the maximum pressure $P_{\rm max}$ and its time-average 
$\bar{P}_{\rm max}$. The time course of $P_{\rm max}$ is shown in 
Fig.~\ref{fig:flow}(c), in which the apparent discontinuities 
arise from abrupt changes in the location of the maxima due to the
random flagellar phases.

\begin{figure}[t]
    \includegraphics[width=0.95\columnwidth]{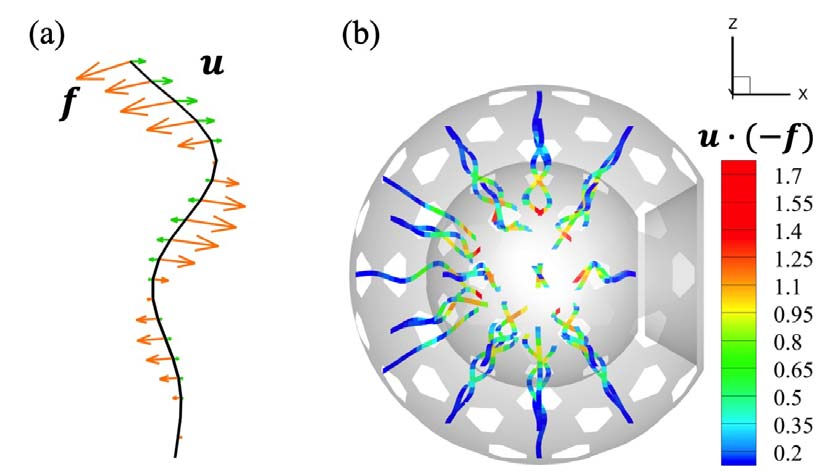}
    \caption{Work rate of flagella. (a) Force density and velocity at node points of a flagellum. (b) Magnitude of the inner product of the force density and the velocity.}
    \label{fig:w}
\end{figure}

\textit{Flagellar force:}
The force density $\bm{f}$ and the velocity $\bm{u}$ of node points on a flagellum 
are shown in Fig.~\ref{fig:w}(a).  These can be used to determine the rate of
working $W^*$ of the entire flagellum in moving fluid as 
\begin{equation}
    W^* = \sum_{i}^{N}\int\bm{u} \cdot (-\bm{f})ds_i.
\label{eq:w}
\end{equation}
With the force density $\bm{f}\sim \mu\bm{u}$ and speeds scaling as $L/T$, we define the 
dimensionless
rate as
\begin{equation}
    W=\frac{W^* T^2}{\mu L^3}.
\end{equation}
Intuitively, we see that the magnitude of $\bm{u}\cdot (-\bm{f})$ of each flagellum is largest near the tip,
as in Fig.~\ref{fig:w}(b), which creates high pressure in the center of the chamber.

\subsection{Output increase with number of flagella}
To clarify the relationship between the number of flagella in a chamber and its
fluid dynamical properties, pumping functions were investigated as the number 
was varied from $10$ to $347$.
The pumping function was evaluated according to the outlet flow rate 
$Q$ from the apopyles, 
the rate of working of the flagella, the maximum pressure $P_{max}$ 
in the chamber (assuming zero pressure at infinity) and 
the mechanical pumping efficiency $\eta$ averaged over time during a 
beat cycle of the 
flagellar, where
\begin{equation}
    \eta = \frac{\bar{Q}\bar{P}_{max}}{\bar W},
\label{eq:eta}
\end{equation}
where overbars indicate time-averaged quantities.

\begin{figure}[t]
    \includegraphics[width=0.98\columnwidth]{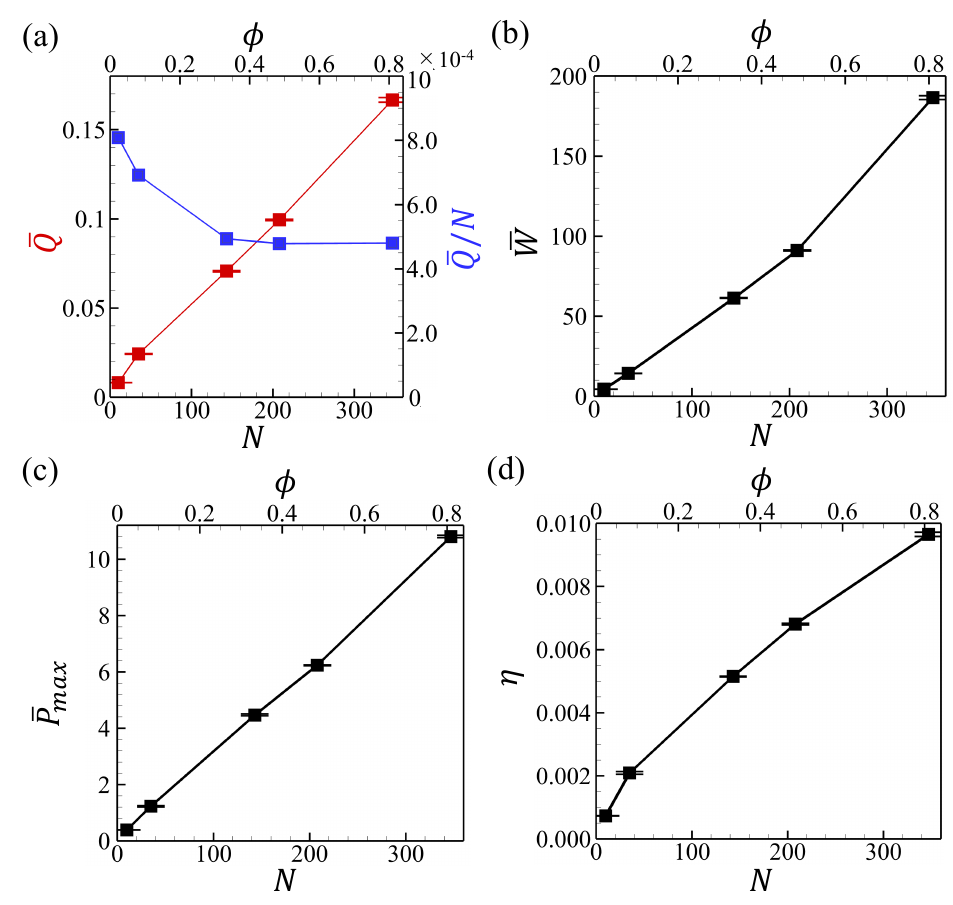}
    \caption{Correlation of pumping functions with the number of flagella. (a) Outlet flow rate. (b) Work rate done by flagella. (c) Maximum pressure in the chamber. (d) Mechanical pumping efficiency.}
    \label{fig:nfla}
\end{figure}

\begin{figure}[t]
    \includegraphics[width=0.98\columnwidth]{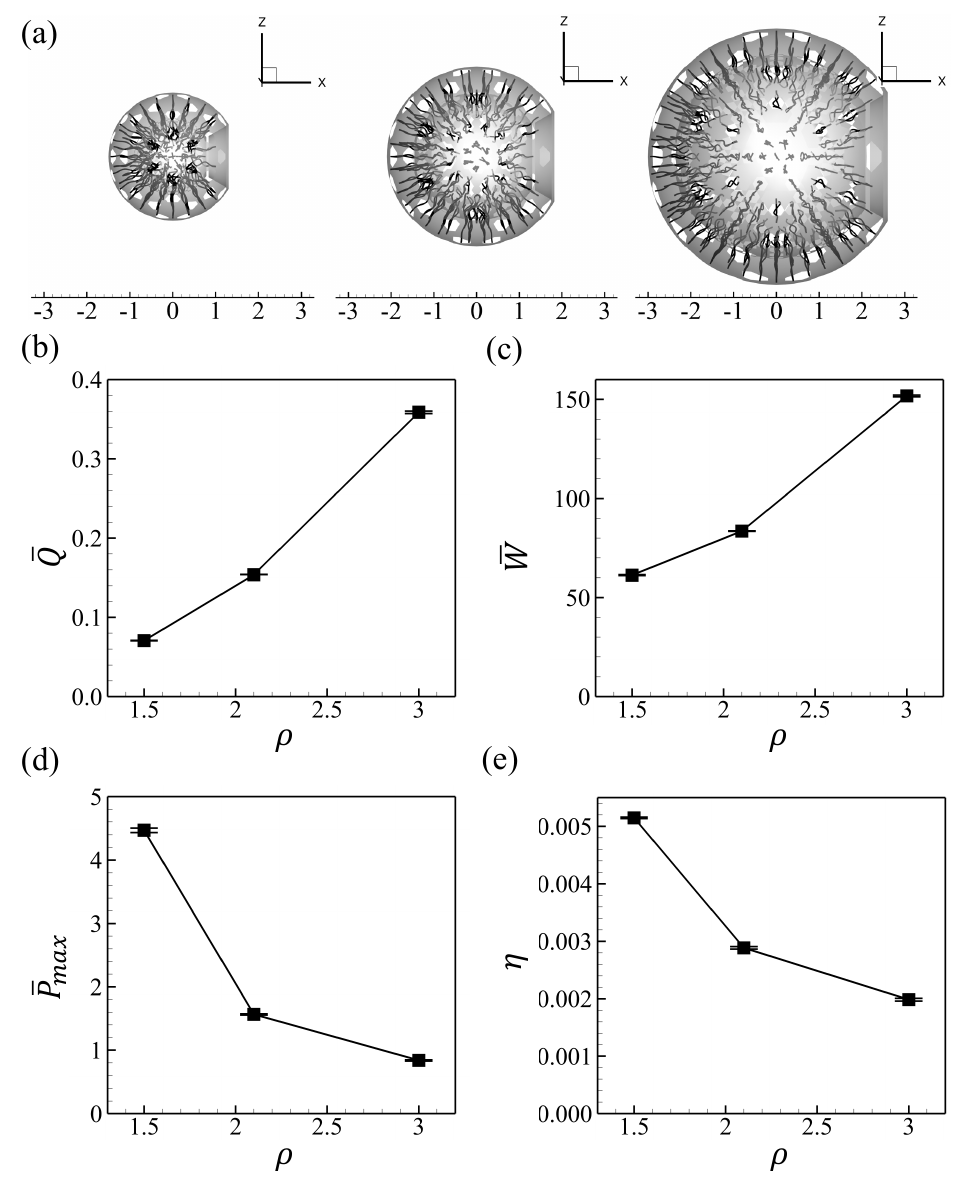}
    \caption{Correlation of pumping function with chamber radius. The length and number density of flagella are fixed. (a) Outlet flow rate. (b) Work rate done by flagella. (c) Maximum pressure in the chamber. (d) Mechanical pumping efficiency.}
    \label{fig:r}
\end{figure}

\begin{figure}
    \includegraphics[width=8.6cm]{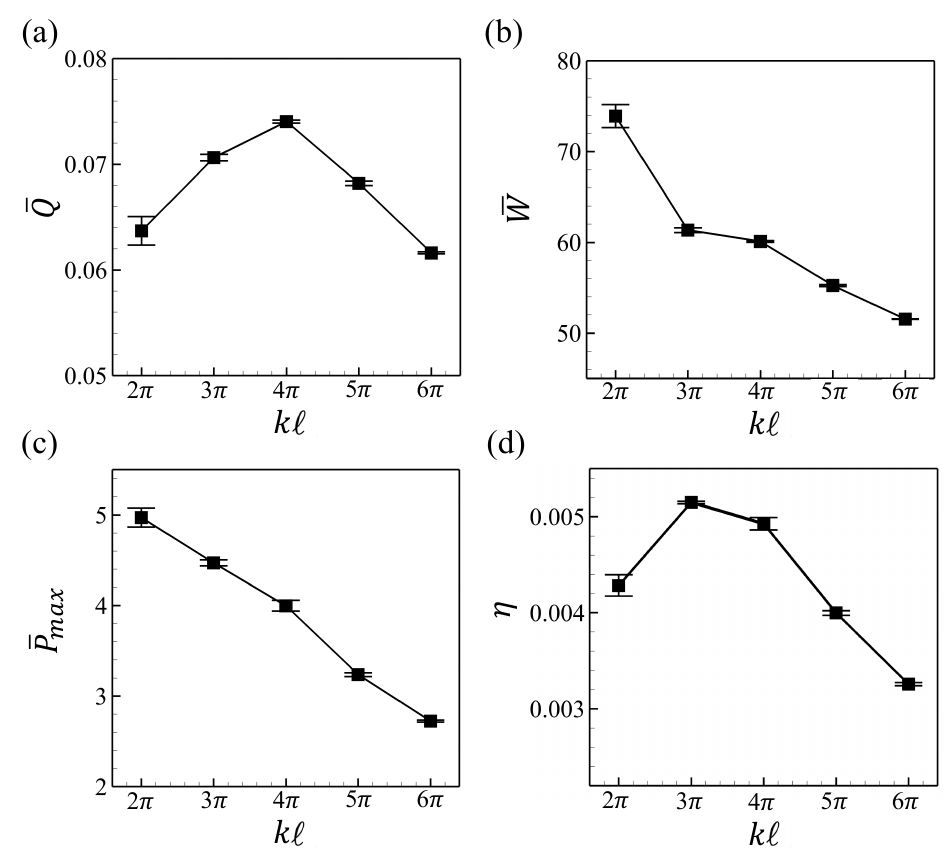}
    \caption{Correlation between the pumping function and wave number of flagella. (a) Outlet flow rate. (b) Work rate done by flagella. (c) Maximum pressure in the chamber. (d) Mechanical pumping efficiency.}
    \label{fig:wn}
\end{figure}

In Fig.~\ref{fig:nfla}, ensemble averaged $\bar{Q}$, $\bar{W}$, $\bar{P}_{max}$, 
and $\eta$ are shown with three independent 
phases and rotation angles of the flagellar beating planes.
The larger the number of flagella (and the greater the packing fraction $\phi$), 
the greater $\bar{Q}$, $\bar{W}$, $\bar{P}_{max}$, 
and $\eta$, indicating better performance as a pump.
Most notably, the average flux $\bar{Q}$ is very nearly linear in $N$ and the
flux per flagellum $\bar{Q}/N$ saturates to a constant value of $5\times 10^{-4}$ at large $N$. 
In interpreting this value it is useful to adopt a simplified view of the flagellum as 
a localized point force acting on the fluid, a model that has experimental support from
studies of {\it Chlamydomonas reinhardtii} \cite{Brumley2014}.  The associated Stokeslet flow field,
if examined in free space without boundaries, has an infinite flux through any plane orthogonal
to the direction of the point force, while for a force orthogonal to a nearby no-slip surface
the flux vanishes by fluid conservation.  This latter result arises from compensating flows away
from the surface near the singularity and toward the surface far away from it.  A numerical computation
of the flux associated only with the outgoing flows driven by a single model flagellum used
in the present calculations, attached to a no-slip wall, gives a value of $\sim 0.01$, a factor 
of $20$ greater than the limit seen in Fig.~\ref{fig:nfla}(a).  While the porosity of the wall
would tend to increase the flux, the tight confinement within the chamber and the typically small
apopyle angle clearly leads to significant cancellation.
For a given chamber radius, 
excluded volume effects from cells and their collars 
limits the number of choanocytes that can be spherically aligned, a limit
that depends on the size and shape of the choanocyte.
The choanocyte diameter varies among species: $3\,\mu$m for 
\textit{Tetilla serica} \cite{Watanabe1978}, $3.7-5\,\mu$m with pseudocylindrical 
shape for 
\textit{Halisarca dujardini} \cite{Gonobobleva2009}, $6\,\mu$m for 
\textit{E. fluviatilis} \cite{Funayama2005} and $5\,\mu$m for \textit{E. muelleri} \cite{Nikko2001}.

\subsection{A smaller chamber radius increases efficiency}
We investigated the effect of the chamber radius on the pumping functions 
such as the efficiency by varying the radius from $1.5L$ to $3.0L$ 
while keeping the number density of flagella fixed.
As shown in Fig.~\ref{fig:r}, the flow rate and work rate increased with the chamber radius as the number of flagella also increased.
On the other hand, the maximum pressure decreased as the chamber radius increased.
The pressure drop was due to the large center space within the chamber, where
there are no direct flagella forces.
As a result, the mechanical pumping efficiency decreased with increasing  radius.
These results indicate that a larger chamber has no efficiency advantage, but rather that it is advantageous to pack more choanocytes into a smaller chamber.

\subsection{Efficiency is maximized at intermediate $k\ell$}

\begin{figure}
\includegraphics[width=8.6cm]{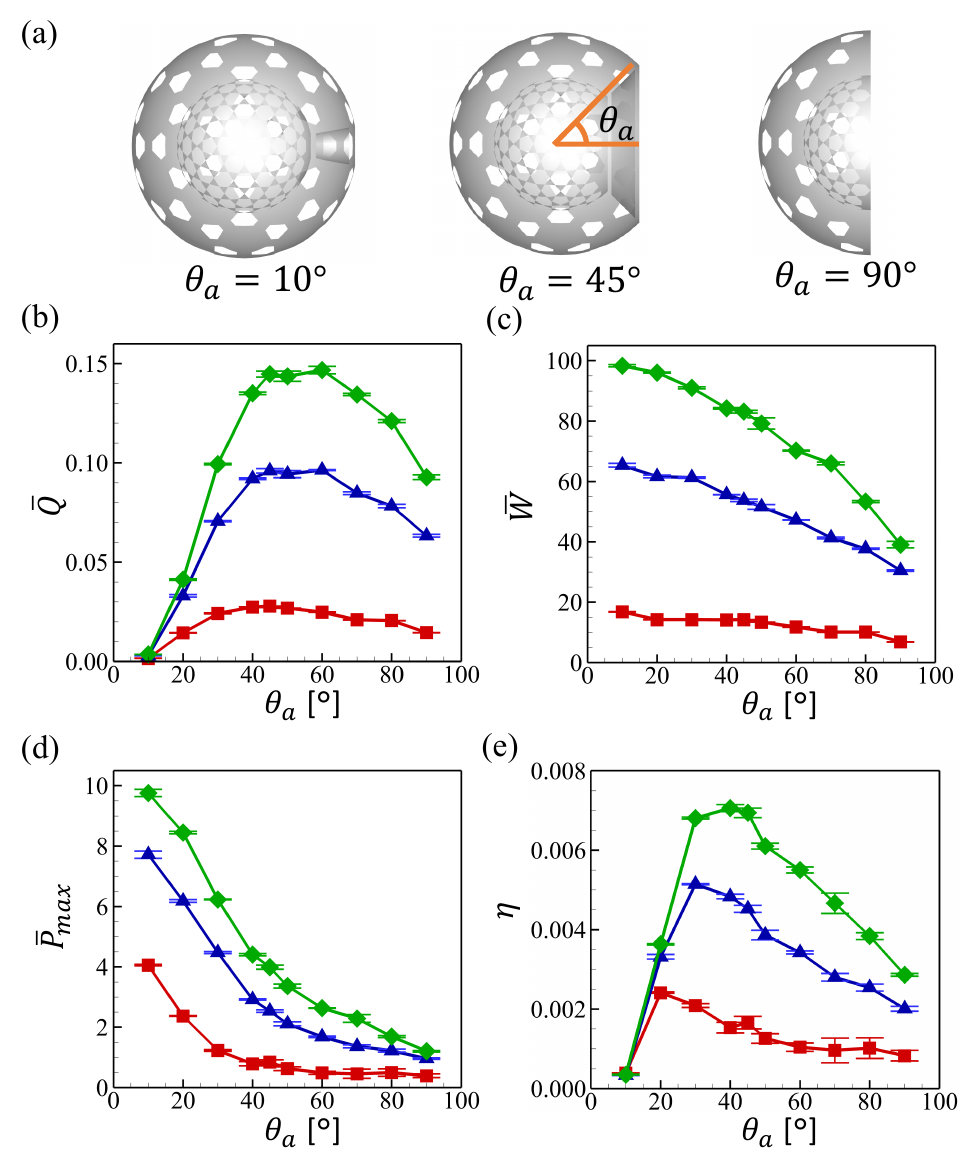}
\caption{Correlation between pumping function and outlet opening angle. 
(a) Computational domain for three values of $\theta_a$.  (b-e) Chamber properties
as a function of opening angle, keeping area fraction $\phi$ in narrow ranges:
red: $0.074-0.093$, blue: $0.31-0.34$, 
and green: $0.39-0.49$. (b) Outlet flow rate. 
(c) Work rate of flagella. (d) Maximum pressure in the chamber. 
(e) Mechanical pumping efficiency.}
\label{fig:theta}
\end{figure}

We studied the effect of changing the flagellar wave number while keeping the
flagellar length fixed. 
Interestingly, as shown in Fig.~\ref{fig:wn}, the outlet flow rate and 
mechanical pumping efficiency exhibit peaks at intermediate values of $k$,
while the particular value of the peak $k$ differs between the two.
The mechanical pumping efficiency reaches a maximum when at the 
relatively low wave number $k\ell=3\pi$, where $\ell$ is projected flagellar length.  
That higher wave numbers lead to reduced efficiency, despite the reduced work 
rate of the flagella, arises from an effect similar to that found when the chamber radius
is reduced;
Since $L$ is fixed, $\ell$ shrinks at higher wave numbers,
increasing the space at the chamber center from which flagella are
absent, and
the central pressure reduces with higher wave number.
The wave number associated with the efficiency is maximized is compared with the experimental results in Section \ref{Comp_wn}.

\subsection{Efficiency is maximized at intermediate $\theta_a$}

While the choanocyte chamber diameter, apopyle area, and diameter were examined in 
previous studies \cite{Larsen1994,Johnston1982,Lays2011}, it has remained 
unclear how the apopyle's aperture ratio affects the pumping function of the chamber.
We examined the pumping function $\theta_a \in [10^\circ,90^\circ]$ while keeping 
the area fraction $\phi$ within each of three narrow ranges, 
adjusting the number of flagella with $\theta_a$ accordingly.
Figure~\ref{fig:theta}(b-e) shows the variation in pumping functions
with $\theta_a$.
With all three flagella densities, similar trends were observed, with small 
variations in the position of various peaks.
The outlet flow rates reaches a maximum at $\theta_a \sim 40^\circ-60^\circ$.
On the other hand, the maximum pressure decreases monotonically 
with $\theta_a$, an effect that arises from the reduction in the 
the number of flagella directed against the bulk flow as $\theta_a$ increases. 
Thus, the spherical shape of the choanocyte chamber has the effect of increasing 
pressure.
Since the pumping efficiency is the product of the maximum pressure and the flow 
rate, its peak in Fig.~\ref{fig:theta}(e) shifts to the lower $\theta_a$ regime of 
$20^\circ-50^\circ$ compared to the case of the flow rate 
(cf. Fig.~\ref{fig:theta}(b)).
From these results, we conclude that flagella around apopyles, which 
seem to disturb unidirectional flow, contribute to creating the high pressure 
rise and that the choanocytes with intermediate but small $\theta_a$ can achieve high mechanical 
pumping efficiency.

\begin{figure}[t]
    \includegraphics[width=0.80\columnwidth]{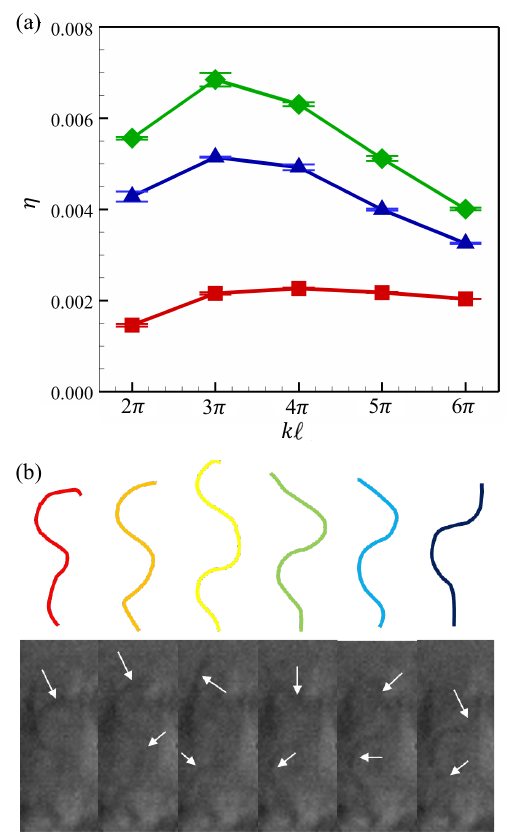}
    \caption{Wave number of flagella. (a) Mechanical pumping efficiency as a function of the wave number with color scheme as in Fig.~\ref{fig:theta} for 
    $\phi$ in the ranges $0.074-0.093$ (red), $0.31-0.34$ (blue), and $0.39-0.49$ (green).
    (b) Flagellar motion of a choanocyte of \textit{E. muelleri}. White arrows indicate flagella. Colored curves represent extracted flagellar waveforms.}
    \label{fig:wave}
\end{figure}

\section{Comparison with experiment}
\label{Comparison}

\begin{table*}
\caption{\label{table:theta}
Dimensions of choanocyte chambers.
}
\begin{ruledtabular}
\begin{tabular}{cccccc}
     &
     \begin{tabular}{c}
     Chamber diameter \\ ($\mathrm{\mu m}$)\\
     \end{tabular} 
     &
     \begin{tabular}{c}
     Apopyle diameter \\ ($\mathrm{\mu m}$)\\ 
     \end{tabular} 
     & 
     \begin{tabular}{c}
     Apopyle area \\ ($\mathrm{\mu m^2}$) \\
     \end{tabular} 
     & $\theta_a$ ($^\circ$) & 
     \begin{tabular}{c}
     Choanocytes \\ per chamber \\
     \end{tabular} \\
     1. \textit{Ephydatia muelleri} \footnote{$n =5$, $N = 3$} & 34.7$\pm$5.0 & 18.1$\pm$4.3 & & 32.2$\pm$6.7 & 112$\pm$31\\
     2. \textit{Haliclona urceolus}\footnote{Data from Larsen et al. \cite{Larsen1994}.} & 30 & 14 & & 27.8 & 80\\
     3. \textit{Haliclona permollis}\footnote{Data from Ludeman et al. \cite{Ludeman2017}.} & 30 & 14 & & 27.8 & 95\\
     4. \textit{Aphrocallistes vastus}\footnotemark[3] & 56 & 26 & & 27.7 & 260\\
     5. \textit{Neopetrosia problematica}\footnotemark[3] & 23.3 & 16.0 & & 43.4 & 80\\
     6. \textit{Haliclona mollis}\footnotemark[3] & 28.5 & 14.1 & & 29.7 & 139\\
     7. \textit{Tethya californiana}\footnotemark[3] & 21.1 & 0.90 & & 2.44 & 99\\
     8. \textit{Callyspongia vaginalis}\footnotemark[3] & 19.7 & 5.97 & & 17.6 & 93\\
     9. \textit{Cliona delitrix}\footnotemark[3] & 16.0 & 4.23 & & 15.3 & 50\\
     10. \textit{Amorphinopsis foetida}\footnote{Data from Dahihande et al. \cite{Dahihande2021}.} & 17.46$\pm$0.13 & & 124.11 & 46.1 & 84.11$\pm$3.02\\
     11. \textit{Calyspongia} sp.\footnotemark[4] & 19.35$\pm$0.2& & 168.54 & 49.2 & 121.11$\pm$4.33\\
     12. \textit{Haliclona} sp.\footnotemark[4] & 19.69$\pm$0.23 & & 113.21 & 37.6 & 68.83$\pm$2.82\\
     13. \textit{Ircinia fusca}\footnotemark[4] & 30.68$\pm$0.29 & & 211.25 & 32.3 & 120.35$\pm$8.98\\
\end{tabular}
\end{ruledtabular}
\end{table*}

\textit{Pumping flow rate:}
\label{Comp_flow}
The numerical results in Section \ref{flow_field}, with parameters taken from 
experiment, yielded time-averaged flow rates of a choanocyte chamber 
of $\bar{Q}\simeq 4\times 10^3 \mathrm{\mu m^3/s}$. This can be compared with 
previous experimental results. The flow rate of the choanocyte chamber of \textit{Haliclona urceolus} was estimated as $Q=4.5\times10^3 \mathrm{\mu m^3/s}$ \cite{Larsen1994}, which is very close to the present results.
In addition, the flow rate per choanocyte of a syconoid type sponge was estimated at 
$48\,\mu$m$^3$/s \cite{Asadzadeh2020}, and if the flow rate scales
with the number of choanocytes, estimated to be $100$ (cf. Table~\ref{table1}), we 
obtain $Q=4.8 \times 10^3\mu$m$^{3}$/s, 
also in good agreement with our results.
Thus, the flow rate obtained in this study is consistent with former experimental observations.

\textit{Flagellar wave number:}
\label{Comp_wn}
Next, we compare the wave number at which the efficiency is maximized with our 
experimental results. The mechanical pumping efficiency as a function of the wave 
number with three different number densities of flagella is shown in Fig.~\ref{fig:wave}(a). 
We see peaks in efficiency with the $k\ell\sim (3-4)\pi$, independent of the number 
density of flagella.
In our experiment, the flagellar motion of \textit{E. muelleri} choanocyte was 
extracted from the high-speed imaging as shown in Fig.~\ref{fig:wave}(b).
From the projected flagellar length $\ell = 9.83\pm 0.74\,\mu$m and the 
wavelength $\lambda=5.98\pm 0.48\, \mu$m, we find $k\ell =(3.30\pm 0.27)\pi$,
in good agreement with the numerical results for the maximum efficiency.
Similar trends are seen in other species. In the case of \textit{S. lacustris}, 
for example, $k \ell \sim 4\pi$ using $\ell =10.4\,\mu$m and $\lambda=5\,\mu$m 
\cite{Asadzadeh2019,Mah2014}.
Thus, the flagellar wave number appears to be tuned to maximize the mechanical pumping efficiency.

\textit{Chamber size and choanocyte density:}
Our experimental observations show that the number of flagella in a choanocyte 
chamber of \textit{E. muelleri} is $\sim 100$, similar to the value $\sim 80$ 
found in earlier work
on choanocytes of \textit{H. urceolus}, in which the chamber diameter 
is $\sim 30\,\mu$m \cite{Larsen1994}.
In several marine sponges, the chamber diameters ranges from $16-31\,\mu$m 
and the number of choanocytes ranges from $32-130$, which indicates that choanocytes are densely packed in the chamber.
There is a positive correlation between the number of choanocytes per chamber and the pumping rate of a sponge according to experimental observations \cite{Dahihande2021}. Hence, it is inferred that the chamber is filled almost completely with as many choanocytes as allowed by the chamber size.
These tendencies are consistent with our finding that larger number of flagella enhances pumping function and efficiency.

\begin{figure}[t]
\includegraphics[width=0.95\columnwidth]{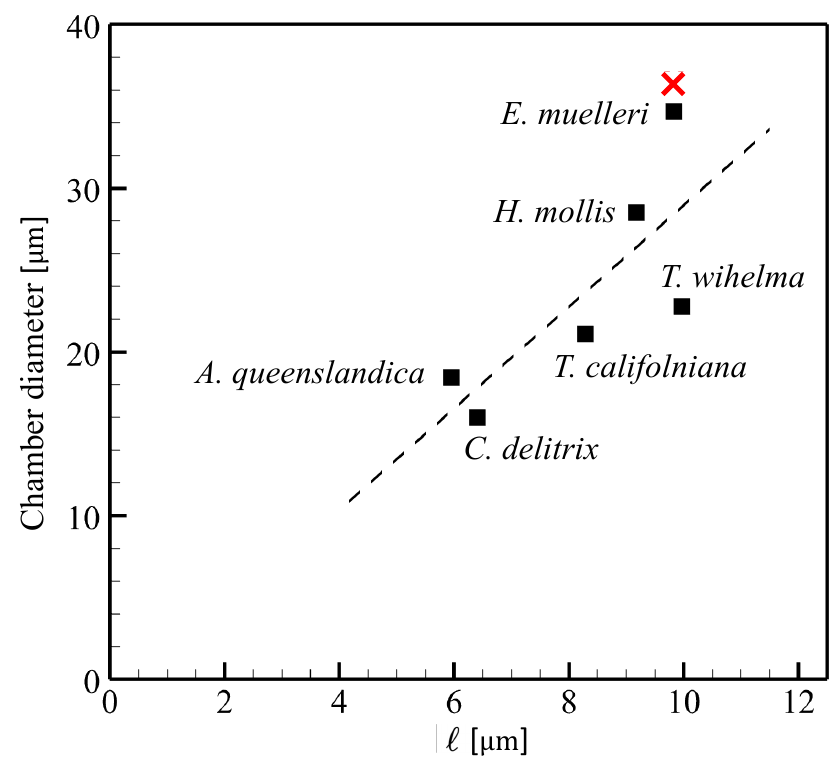}
\caption{Correlation between the projected flagellar length and the choanocyte 
chamber diameter. If not specified, the values were measured manually from 
images; \textit{Tethya wilhelma} from Hammel et al. \cite{Hammel2014}; 
\textit{Haliclona mollis}, \textit{Tathya californiana}, and \textit{Cliona 
delitrix} from Ludeman et al. \cite{Ludeman2017}; \textit{Amphimedon 
queenslandica} from Sogabe et al. \cite{Sogabe2016}. \red{$\times$} 
indicates the hypothetical minimum diameter of \textit{E. muelleri} 
that is obtained under the condition that adjacent flagella do not touch.}
\label{fig:l}
\end{figure}

As noted earlier, there is an upper limit on the number density of choanocytes due to the excluded volume of cells and the aperture ratio of apopyles and prosopyles.
Hence, to increase the number of choanocytes, the choanocyte chamber is forced to make its radius larger. On the other hand, when the radius is large, the pumping efficiency becomes low. These two conditions are contradictory, so a balance 
must be reached between them.
To find balanced conditions in nature, we plotted the correlation of the flagellar length and the choanocyte chamber diameter for various species of sponge as shown in Fig.~\ref{fig:l}.
We see an obvious positive correlation between them, indicating that the chamber 
diameter increases with flagellar length.
We analytically calculated the hypothetical minimum diameter of \textit{E. muelleri} under the condition that adjacent flagella of amplitude 3.16 $\mathrm{\mu m}$ do not contact each other. The result is 36.4 $\mathrm{\mu m}$, which is close to the actual diameter of 34.7 $\mathrm{\mu m}$.
This implies that the chamber diameter is designed to be as small as possible while avoiding overlapping flagella.
Smaller chamber diameters may have other advantages, such as the ability to accommodate a larger number of cells in a smaller volume and greater flexibility in the design of the canal in the entire of sponge body.

\textit{Outlet opening angle:}
\label{Comp_oa}
Finally, we compare the experimentally observed opening angles with the range 
$\theta_a = 20^\circ -50^\circ$ found to be maximal for pumping efficiency. We observed several cross-sections of a choanocyte 
chamber of \textit{E. muelleri} by shifting the focal plane as shown in 
Fig.~\ref{fig1}(c-f).
From the image around the apopyle (cf. the top focal plane in Fig.~\ref{fig1}(e)), 
we measure the apopyle diameter.  Using this value and the 
chamber diameter, and assuming a spherical 
chamber shape, the average apopyle opening angle was found to be 
be $\theta_a=32.2^\circ$.
In the case of \textit{E. muelleri}, the number density of flagella is $0.032\mathrm{/\mu m^2}$,
which, with the chamber diameter of $34.7 \mathrm{\mu m}$, and the flagella number of $112$
gives a coverage fraction {\color{blue}$\phi=0.25$}, associated with the blue data in 
Fig.~\ref{fig:theta}(e) for pumping efficiency, which is maximized at 
$\theta =30^\circ$.  The very 
good agreement with the observed value for \textit{E. muelleri} 
indicates that the natural configuration of the choanocyte chamber 
of \textit{E. muelleri} optimizes the mechanical pumping efficiency.

\begin{figure}[t]
    \includegraphics[width=8.4cm]{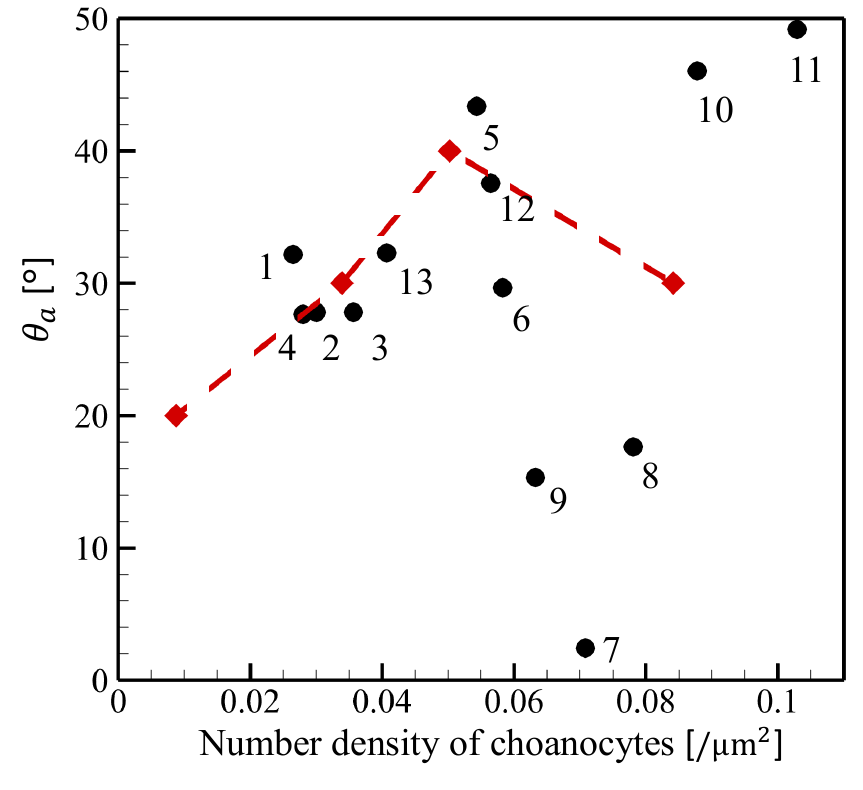}
    \caption{Correlation of the opening angles of apopyles and the number density of choanocyte. 
    Black symbols and numbers correspond to the species listed in Table~\ref{table:theta}. Red 
    symbols indicate the angle at which the numerically obtained mechanical pumping efficiency 
    reaches its maximum at each of four number densities used in the simulation.}
    \label{fig:density}
\end{figure}

The geometric properties of choanocyte chambers in many sponge species 
that have been reported previously are collected in Table~\ref{table:theta}. To compare the 
present study with these former ones, we plotted the opening angle $\theta_a$ versus the 
number density of choanocytes in a chamber. 
In estimating these values, we assumed the chamber is spherical. 
Studies on \textit{E. muelleri}, \textit{Haliclona urceolus}, \textit{Haliclona permollis}, \textit{Aphocallistes vastus}, \textit{Neopetrosia problematica}, \textit{Haliclona mollis}, \textit{Tethya californiana}, \textit{Callyspongia vaginalis}, and \textit{Cliona delitrix} \cite{Larsen1994,Ludeman2017}
explicitly reported information about apopyles, so we could fully incorporate the effect of apopyles when calculating the number density of choanocytes. In studies on \textit{Amorphinopsis foetida}, \textit{Callyspongia} sp., \textit{Haliclona} sp. and \textit{Ircinia fusca} \cite{Dahihande2021},
on the other hand, the apopyle area was not reported and we calculated the number density without taking apopyles into account, thus underestimating the number density. 
The results are plotted in Fig.~\ref{fig:density}, in which the present results on those
values of $\theta_a$ that correspond to the maximum mechanical pumping efficiency at various
densities are indicated in red.
The opening angles of natural sponge choanocyte chambers are clearly close to the 
computational results for those with maximum efficiency, thus demonstrating that 
many species of sponge have choanocyte chambers that, like \textit{E. muelleri}, 
can achieve high mechanical pumping efficiency.

\section{Point Force Model}
\label{Force_model}

A recurring theme in biological fluid dynamics is the representation of flow fields around 
multiflagellated organisms by suitable singularities in Stokes flow.  For example, 
experimental studies of 
freely-swimming colonies of the green alga \textit{Volvox carteri}, a spheroid consisting of 
$\sim 10^3$ biflagellated cells on its surface, have shown a dominant far-field behavior
associated with a Stokeslet arising from the density offset between the colony and the surrounding
water; a single point force accurately summarizes the effects of a thousand cilia \cite{Polin2010}. 
At smaller scales, an accurate representation of the swirling flows near a single biflagellated alga \textit{Chlamydomonas reinhardtii} when it swims in a breaststroke fashion requires three
point forces: one for the cell body and one each for the opposing flagella \cite{Polin2010,Ishikawa2024}.

Returning to the densely-packing choanocyte chambers, it is natural to examine the extent to which
the fluid dynamical properties we have found in the numerical studies described above 
can be represented by the action of one or several point forces.  Such a representation
would be useful in understanding the input-output characteristics of the chamber, particularly 
within a coarse-grained model of the the sponge network.

\begin{figure}[t]
    \includegraphics[width=0.85\columnwidth]{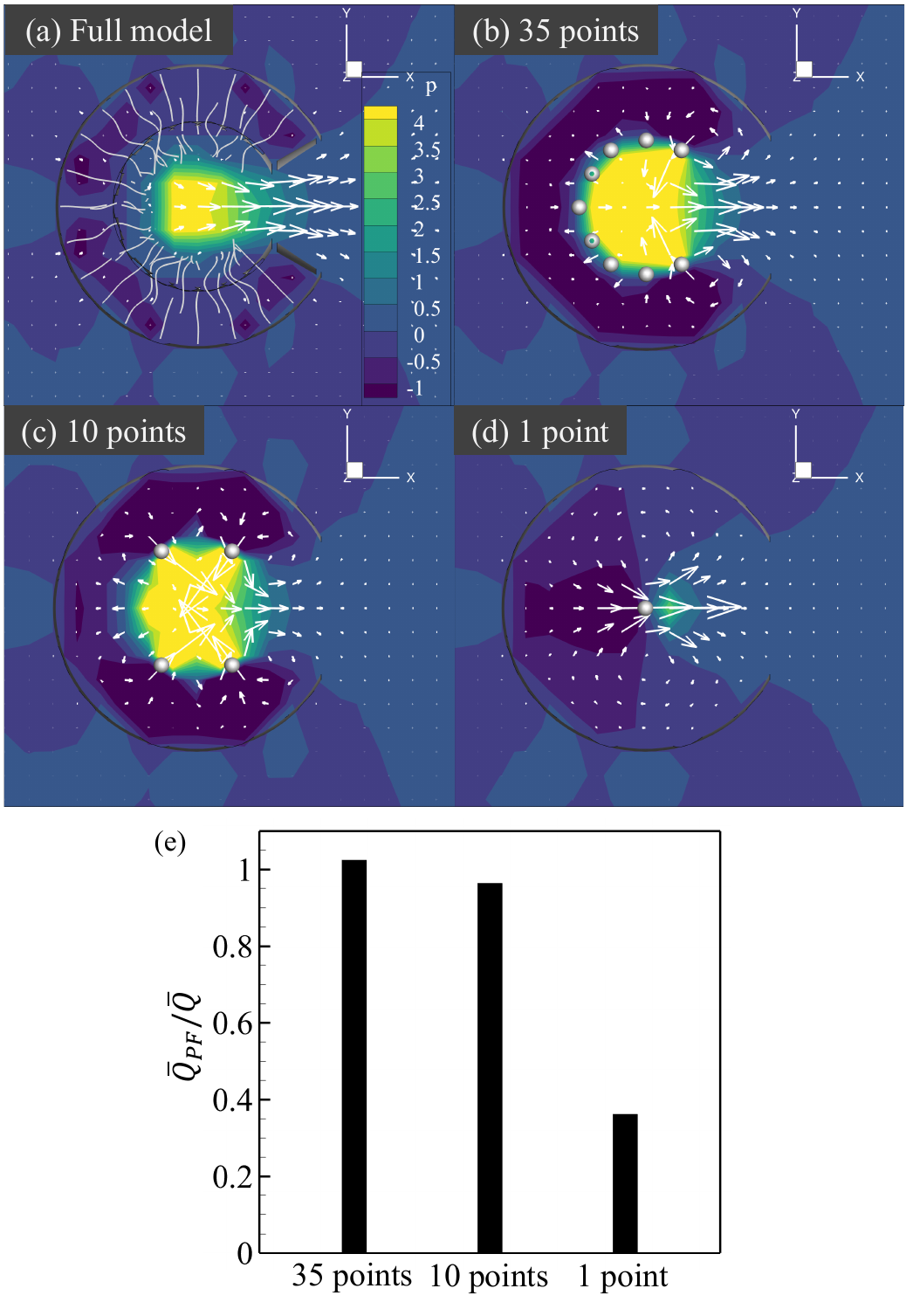}
    \caption{Flow generated by coarse-grained models of a choanocyte chamber. (a-d) Flow and pressure fields generated by the full simulation (a), 35 point forces (b), 10 point forces (c), and a single point force (d), where dots indicate the positions of the point forces.
    (e) Flow rate of the coarse-grained models $\bar{Q}_{PF}$ relative to the flow rate of the full simulation $\bar{Q}$.}
    \label{fig:coarse}
\end{figure}

In the original full simulation, there were flagella, the reticulum and the cone cell ring 
inside a spherical chamber, and the flow was generated as shown in Fig.~\ref{fig:coarse}(a). 
We then integrated the forces acting on these internal structures consolidated them into 
a set of point forces.
Fig.~\ref{fig:coarse}(b-d) shows the flow field when the internal structure is divided 
almost equally into $M$ sections and $M$ point forces are applied at the indicated locations, for
the cases (b) $M=35$, (c) $10$ and (d) $1$ point force at the sphere center.
In all of these coarse-grained models, a flow can be observed from the center of the spherical chamber towards the apopyle.

Fig.~\ref{fig:coarse}(a-d) also show the pressure field. 
Due to the different distribution of forces, the pressure distribution is also altered  
from one value of $M$ to another, but for all values $M>1$ there is a similar scale of the high pressure
acting at the centre of the chamber.  It is only when $M=1$ that there is a significant reduction
in the central pressure.  This can be seen in the flow rate
of the coarse-grained models relative to that in the fully-resolved numerics, shown in 
Fig.~\ref{fig:coarse}(e).  Even for $M$ as small as $10$ flux is quantitatively matched within 
a few percent.  But the single point force drastically underestimates the flux as a direct 
consequence of underestimating the central pressure.  We conclude that the effects of flagella
around the apopyle, opposed to the directional flow, are indeed crucial to obtain the scale of
flux from choanocyte chambers seen in experiment.  At the same time, a significantly simplified 
representation of the chamber is possible.

\section{Conclusions}
\label{Conclusions}
In this study, we combined direct imaging of choanocyte chambers in living sponges with computational studies of many-flagellum models of their fluid mechanics to unravel the biological significance of the spherical shape of choanocyte chambers. 
We determined that there are ideal conditions for achieving maximum mechanical pumping efficiency for such a geometry.
First, to achieve better efficiency, the chamber radius should be as small as possible and the number density of choanocytes per chamber should be as large as possible.
Excluded volume effects limit the range of densities, and 
we found for \textit{E. muelleri} and several other species that the chamber diameter is as small as possible while avoiding overlapping flagella.
Second, the optimum wave number for mechanical pumping efficiency was found to be 
$k\ell\sim 3\pi$ when the chamber radius is $1.5$ times the flagellar length. 
This linkage was in good agreement with experimental results on \textit{E. muelleri}.
Third, the optimum opening angle for mechanical pumping efficiency 
was found to be in the range $20^\circ - 50^\circ$, with a value for 
\textit{E. muelleri} of $30^\circ$, which was again in good agreement with 
the experimental result of $32.2^\circ$. The computational estimate 
of the optimum opening angle is also in good agreement with those of many other species.

In addition, our computational analysis revealed that those flagella that beat against
the dominant flow play a role in raising the pressure inside the choanocyte chamber. 
As a result, the mechanical pumping efficiency\textemdash calculated from the 
pressure rise and flow rate\textemdash reaches a maximum at a modest outlet opening angle.
Given that high pumping efficiency is achieved in many species 
(cf. Fig.~\ref{fig:density}), the choanocyte chamber may have evolved
to optimize this feature, which can be viewed as a means of overcoming the high 
fluid dynamical resistance of the complex canal.
Finally, out investigation of the fidelity of coarse-grained models of the 
chamber reveals that such simplifications can be taken too far to be accurate, 
particularly when much of the phenomena of interest are in the near-field regime,
as they appear to be for sponge choanocyte chambers.  This result does not preclude
appropriate coarse-grained models for networks of chambers, only that the input-output
characteristics are nontrivial.  Logical next steps include development
of such models, and on a more fundamental level understanding the
developmental processes that lead to such complex network architecture in leuconoid
sponges.

\begin{acknowledgments}
This work was supported by JST SPRING, Grant Number JPMJSP2114, JST PRESTO (Grant Number JPMJPR2142).
K.K. was supported by the Japan Society for the Promotion of Science Grant-in-Aid for Scientific Research (21H05303, 21H05306, 22H01394) and  JST FOREST Program, Grant Number JPMJFR2024.
T.I. was supported by the Japan Society for the Promotion of Science Grant-in-Aid for Scientific Research (JSPS KAKENHI Grant No. 21H04999 and 21H05308). R.E.G. acknowledges support from the Wellcome Trust Investigator Grant 207510/Z/17/Z and The John Templeton Foundation.
\end{acknowledgments}

\bibliography{reference}

\end{document}